\documentclass[aps,prb,superscriptaddress,twocolumn,floatfix]{revtex4}
\usepackage{color,graphicx,pstricks}
\usepackage{psfrag}
\usepackage{textcomp}
\usepackage{hyperref}
\hypersetup{
  colorlinks=true,
  citecolor=blue,
  linkcolor=magenta,
  }
\usepackage{amsmath}
\begin{document}

\title {Layer Resolved Magnetotransport Properties in Antiferromagnetic/Paramagnetic Superlattices}

\author{Sandip Halder}
\affiliation{Theory Division, Saha Institute of Nuclear Physics,
  A CI of Homi Bhabha National Institute, Kolkata-700064, India}
\author{Sourav Chakraborty}
\affiliation{Theory Division, Saha Institute of Nuclear Physics,
  A CI of Homi Bhabha National Institute, Kolkata-700064, India}
\affiliation{ {Department of Condensed Matter and Materials Physics,
S. N. Bose National Centre for Basic Sciences, JD Block, Sector III, Salt Lake, Kolkata-700106, India}}
\author{Kalpataru Pradhan}
\affiliation{Theory Division, Saha Institute of Nuclear Physics,
  A CI of Homi Bhabha National Institute, Kolkata-700064, India}

\begin{abstract}
We investigate the layer resolved magnetotransport properties of the
antiferromagnetic/paramagnetic superlattices based on one band half-filled
Hubbard model in three dimensions. In our set up the correlated layers
(with on-site repulsion strength $U \ne$ 0) are intercalated between the
uncorrelated (U = 0) layers. Our calculations based on the semi-classical
Monte-Carlo technique show that the magnetic moments are induced in the
uncorrelated layers at low temperatures due to kinetic hopping of the
carriers across the interface. The average induced magnetic moment in
the uncorrelated layer varies nonmonotonically with the $U$ values of the
correlated layer. Interestingly, the induced magnetic moments are
antiferromagnetically arranged in uncorrelated layers and mediates the
antiferromagnetic ordering between correlated layers. As a result the
whole SL system turns out to be antiferromagnetic insulating at low
temperatures. For $U \sim$ bandwidth the local moments in the
correlated planes increases as a function of the distance from the interface.
Expectedly our in-plane resistivity calculations show that the metal
insulator transition temperature of the central plane is larger than
the edge planes in the correlated layers. On the other hand, although
the induced moments in uncorrelated planes decreases considerably as
move from edge planes to center planes the metal insulator transition
temperature remains more or less same for all planes. The induced moments
in uncorrelated layers gradually dissipates with increasing the thickness
of uncorrelated layer and as a result the long range antiferromagnetic
ordering vanishes in the superlattices similar to the experiments. 
\end{abstract}

\maketitle

\section{Introduction}
Correlated transition metal oxide heterostructures are extensively studied
nowadays due to their unusual intriguing interfacial
properties~\cite{Hwang,Chakhalian,Chakhalian1,Bhattacharya,Zubko,Schlom,Middey,Stahn,Chakhalian2,Takahashi,Boris}.
Various reconstructions such as electronic~\cite{Okamoto,Okamoto1,Cao,Wang},
magnetic~\cite{Bruno,Lee}, orbitals~\cite{Tebano,Tebano1,Chakhalian3,Peng,Liu,Liu1,Chen}
and structural reconstructions~\cite{Dekker,Boschker,XLi} at the interface
give rise to many fascinating and promising phenomena that are absent in
the constituent bulk materials~\cite{May,Chakhalian2,Qin,Grisolia,Ohtomo,Brinkman,Reyren,Ohtomo1,Zheng,Takamura,Hoffman1,Garcia,Hwang}.
These emerging phenomena like metallicity~\cite{Ohtomo,Nakagawa},
magnetism~\cite{Koida,Brinkman} and superconductivity~\cite{Reyren,Gozer} and
their coexistence~\cite{Dikin,Li,Bert} in the $LaAlO_{3}/SrTiO_{3}$
heterostructures, metallicity in the interfaces of Mott insulator and band
insulator~\cite{Ohtomo1,Seo}, induced magnetism in the paramagnetic
layers in magnetic/paramagnetic superlattices~\cite{Gibert,Gibert1,Lee1,Dong,Hoffman2,Piamonteze,Grutter,Zhou},
and ferromagnetism at the interfaces of paramagnetic/paramagnetic
superlattices~\cite{Zheng} pose many elemental scientific questions that
need to be addressed. Pinning down the underlying mechanism remains a
challenging task for both theory and experiments and remains a subject of
active research~\cite{Mannhart}. A very good knowledge about the structural
and/or the chemical complexity and the modulation of the elementary phases
at the interfaces are crucial to design artificially new functional
materials based on oxide heterostructures~\cite{Bhattacharya,Zubko}.

In $LaTiO_{3}/SrTiO_{3}$ (LTO/STO) superlattices (SL), where LTO is an
antiferromagnetic Mott-insulator and STO is a nonmagnetic band insulator,
an unexpected conducting and magnetic phase appears at the interfaces~\cite{Ohtomo1,Seo}.
As a matter of fact the electrochemical potential difference at the
interface facilitate the leakage of the charge from the filled LTO Ti-$d$
band to the empty STO Ti-$d$ band and promote the magnetic and conducting
state at the interface~\cite{Okamoto}. Interestingly, the amount
of charge transfer across the interface can be controlled by the relative
thickness of the STO layer~\cite{Garcia1}. Induced magnetic moments are also
observed in STO layer in $La_{0.7}Sr_{0.3}MnO_{3}/SrTiO_{3}$ (LSMO/STO)
and $LaMnO_{3}/SrTiO_{3}$ (LMO/STO) SLs~\cite{Bruno,Garcia}.
These induced magnetic moments in Ti controls the overall magnetic and
transport properties of the superlattice system when the thickness of the
STO layer is less than 1 nm~\cite{Bruno}. At the same time the thickness of
LMO layer, due to the orbital reconstructions at the interface of
LMO/STO superlattices~\cite{Garcia,Garcia1}, provides a new pathway to tune
the interaction between the manganite and the titanate~\cite{Garcia}. Density
functional theory studies of LMO/STO superlattices also show that the
magnetic and electronic properties can be controlled by the
thickness of the LMO and STO layers~\cite{Jilili,Cossu}.

Charge transfer from the Mn to the Ni atoms also plays a vital role 
in LMO and LNO ($LaNiO_{3}$) superlattices~\cite{Gibert,Gibert1,Lee1,Dong,Hoffman2,Piamonteze,Grutter}
in inducing magnetic moments in LNO layer. These induced magnetic moments 
drives a metal-insulator transition when the thickness of the LNO layers
reduces to 2 unit cells or less~\cite{Hoffman2,Boris,Chaloupka,Zhou}.
Superlattices comprised of doped Mott-Hubbard insulator $LaTiO_{3+\delta}$
and LNO also offers an unusual electronic structure at the interfaces
facilitated by the charge transfer from $Ti$ to $Ni$ sites~\cite{Cao}.
 Consequently an insulating ground state along with orbital polarization
and $e_{g}$ orbital band splitting like mechanisms are observed at the interface.
Further, it is observed that the metal-insulator transition temperature decreases
as one increases the width of the $LaNiO_{3}$ layer in $NdNiO_{3}/LaNiO_{3}$
SLs~\cite{Nguyen}. On the other hand it is shown that the antiferromagnetic
transition temperature decreases with decreasing the layer thickness of
the antiferromagnetic $CeIn_{3}$ layer in $CeIn_{3}$/$LaIn_{3}$
(antiferromagnetic/metallic) SLs~\cite{Shishido}. In fact, the
antiferromagnetic order vanishes when the thickness of $CeIn_{3}$
reaches as thin as 2 unit cells.

In $NiO/Pd$ superlattices, a type of antiferromagnetic/paramagnetic
superlattices, magnetization is induced in the interfacial metallic $Pd$ layers
due to the proximity effect of the AF $NiO$~\cite{Saha,Manago,Manago1,Manago2,Manago3}.
It is also observed that the induced magnetic moment in $Pd$ layers decreases
with distance from the interfaces and vanishes at a distance $35\AA{}$ away from
interface~\cite{Manago1}. Induced magnetic moments are also observed
in the metallic interfacial $Cu$ layers in $CuO/Cu$ (AF/PM) superlattices~\cite{Munakata}.
Model Hamiltonian based calculations indicate that the magnetic moments are
induced in the metallic layers due to the proximity of the AF insulating
layers~\cite{Jiang,Zujev,Euverte,Mondaini}. In addition, it is
also proposed that the metallicity penetrates in to the insulating AF
layers~\cite{Jiang}, but details of the transport calculations are not provided.
More and more theoretical studies are needed to understand the
magnetotransport properties of magnetic/paramagnetic SLs.

Motivated by the fascinating experimental results in complex oxide
superlattices we investigate the magnetotransport properties of the
correlated/uncorrelated (antiferromagnetic/paramagnetic) superlattices.
In order to explore various phenomena we analyze a range of 3D superlattices
based on one band Hubbard model where correlated ($U \neq 0$) layers of width
$w_{U}$ and the uncorrelated ($U = 0$) layers of width $w_{0}$ are
periodically arranged as shown in Fig.~\ref{sch_sl}. At half-filling our
uncorrelated layers mimics the paramagnetic metallic (PM-M) state where as
the correlated layers imitate the antiferromagnetic insulating state (AF-I)
layers. Overall, our study unveils the key role of mutual cooperation of
the induced magnetic moments in the uncorrelated (paramagnetic) layers
and the local moment in the correlated (antiferromagnetic) layers in
establishing the antiferromagnetic insulating nature of the whole
antiferromagnetic/paramagnetic superlattices. In addition, we emphasize that
the induced moments in uncorrelated planes decreases considerably as a function
of the distance from the interface. As a result the induced moments
completely dissipates with increasing the thickness of uncorrelated
layers and nullifies the antiferromagnetic ordering among correlated layers.
We comprehensively show that the strength of the local moments  and the
metal-insulator transition temperature increases as one moves from the edge
planes to the center plane in correlated layers.

We organize the paper in the following way:
In section {\bf II} we layout the model Hamiltonian associated with
the correlated/uncorrelated superlattices and briefly discuss the numerical
methodology. We outline different physical observables to study magnetotransport
properties of the antiferromagnetic/paramagnetic SLs in section {\bf III}.
Then in section {\bf IV} we present the magnetic and transport properties and
construct the $U-T$ phase diagrams for different $w_{U}$/1 superlattices. We
establish the mutual cooperation between the magnetic ordering between the
correlated and the uncorrelated layers in section {\bf V} and present the plane
resolved magnetic and transport properties of individual layers in section {\bf VI}.
In section {\bf VII} we briefly analyze the 1/$w_{0}$ SLs and then outline
the $x-T$ phase diagram in section {\bf VII}. We thoroughly study 3/3 SL in
section {\bf VII}. At the end, we summarize our results in section {\bf X}.

\section{Model Hamiltonian and Method}
To study of the magnetotransport properties of antiferromagnetic/paramagnetic
superlattices (SLs) we consider following electron-hole symmetric one orbital
Hubbard model:

\begin{align}
H =&-t\sum_{<i,j>,\sigma}(c_{i,\sigma}^{\dagger}c_{j,\sigma}+H.c.)\nonumber \\
&+U\sum_{i}(n_{i,\uparrow}-\frac{1}{2})(n_{i,\downarrow}-\frac{1}{2})\nonumber 
-\mu\sum_{i}c_{i,\sigma}^{\dagger}c_{i,\sigma}\nonumber 
\end{align}

\noindent
where $c_{i,\sigma}^{\dagger}(c_{i,\sigma})$ is the electron creation
(annihilation) operator at site $i$ with spin $\sigma$ ($\uparrow$ or,
$\downarrow$) and $t$ is the hopping amplitude between the nearest neighbors
sites. In the second term $U$ is the strength of on-site Coulomb repulsion between
two electrons of opposite spin at site $i$ and
$n_{i,\sigma} = c_{i,\sigma}^{\dagger} c_{i,\sigma}$ is the spin sum occupation
number operator at site $i$. $\mu$ is the chemical potential which controls the
overall density of the system. In our electron-hole symmetric model Hamiltonian
$\mu = 0$ for half-filling case. Then neglecting the constant term we write down
the Hamiltonian in the following form: 
\begin{align}
H =&-t\sum_{<i,j>,\sigma}(c_{i,\sigma}^{\dagger}c_{j,\sigma}+H.c.)\nonumber \\
&+U\sum_{i}n_{i,\uparrow}n_{i,\downarrow}-\frac{U}{2}\sum_{i}n_{i}\nonumber \\
&=H_{0}+H_{int}\nonumber
\end{align}

\noindent
where $H_{0}$ contains the one body part (i.e. quadratic terms)
and $H_{int}$ denotes the interacting part (i.e. quartic term) of the Hamiltonian.
Now to solve the Hamiltonian using Monte Carlo method the quartic
interaction term can be transformed into combination of two different
quadratic terms to set up the Hubbard-Stratonovich (HS) decomposition:  
\begin{center}
$Un_{i,\uparrow}n_{i,\downarrow}=U\left[\frac{1}{4}n_{i}^{2}
-({\bf{S}}_{i}.\hat{\Omega})^{2})\right]$\\
\end{center}
where ${\bf{S}}_{i} (= \frac{\hbar}{2}\sum_{\alpha,\beta}c_{i,\alpha}^{\dagger}{\bf{\sigma}}_{\alpha,\beta}c_{i,\beta}$
; ${\bf \sigma}$ are the Pauli matrices) is the
spin at $i^{th}$ site and $\hat{\Omega}$ is the arbitrary unit vector.
The partition function of the model Hamiltonian is written as
$Z = Tre^{-\beta (H_{0}+H_{int})}$ where  $\beta = 1/T$ and the Boltzmann
constant $k_{B}$  and $\hbar$ are set to $1$. The interval $[0,\beta]$ is divided into 
$M$ equally spaced slices of width $\Delta\tau$ such that $\beta = M\Delta\tau$.
For large values of $M$ i.e. in the limit $\Delta\tau\rightarrow0$, using
Suzuki-Trotter transformation we can write
$e^{-\beta(H_{0}+H_{int})}=(e^{-\Delta\tau H_{0}}e^{-\Delta\tau H_{int}})^{M}$
up to first order in $\Delta\tau$. After that by implementing the
Hubbard-Stratonovich transformation the interacting part of the partition
function $e^{-\Delta\tau U\sum_{i}[\frac{1}{4}n_{i}^{2}-({\bf{S}}_{i}.\hat{\Omega})^{2}]}$
for a generic  time slice `$l$' can be expressed as 
\begin{align}
\sim\int &d\phi_{i}(l) d\Delta_{i}(l)d^{2}\Omega_{i}(l)\nonumber \\
& \times e^{-\Delta\tau [\sum_{i}\{\frac{{\phi_{i}(l)}^{2}}{2}
+i\phi_{i}(l)n_{i}+\frac{{\Delta_{i}(l)}^{2}}{2}
-2\Delta_{i}(l){\hat{\Omega}}_{i}(l).{\bf{S}}_{i}\}].}\nonumber
\end{align}

The auxiliary field $\phi_{i}(l)$ [$\Delta_{i}(l)$] introduced by the
Hubbard-Stratonovich decomposition is coupled with the charge density $n_{i}$
(spin vector ${\bf{S}}_{i}$). Now we define a new vector auxiliary
field ${\bf{m}}_{i}(l) = \Delta_{i}(l)\hat{\Omega}_{i}(l)$.
Finally we obtain the total partition function as

\begin{align}
Z=const. & \times  Tr \prod_{l=M}^{1} \int d\phi_{i}(l)d^{3}{\bf{m}}_{i}(l)\times \nonumber\\
& e^{-\Delta\tau[H_{0}+\sum_{i}\{\frac{{\phi_{i}(l)}^{2}}{2}
+i\phi_{i}(l)n_{i}+\frac{{{\bf{m}}_{i}(l)}^{2}}{2}
-2{\bf{m}}_{i}.{\bf{S}}_{i}\}].}\nonumber
\end{align}

The product follows the time ordered product, where the time slice
$l$ runs from $M$ to $1$. From the partition function we can extract an
effective Hamiltonian. At this moment we make two major approximations:
(i) by freezing the $\tau$ dependence of the auxiliary fields and retaining
the spatial fluctuations of the auxiliary fields, (ii) using the 
saddle point value of $i\phi_{i}(l) = \frac{U}{2}<n_{i}>$. After all
by redefining ${\bf{m}}_{i}\rightarrow \frac{U}{2}{\bf{m}}_{i}$,
we obtain the following effective Hamiltonian~\cite{Mukherjee,Chakraborty1}:

\begin{align}
H_{eff}=& -t\sum_{<i,j>,\sigma}(c_{i,\sigma}^{\dagger}c_{j,\sigma}+H.c.)\nonumber \\
&+\frac{U}{2}\sum_{i}(<n_{i}>n_{i}-{\bf{m}}_{i}.\sigma_{i})\nonumber \\
& +\frac{U}{4}\sum_{i}({{\bf{m}}_{i}}^{2}-{<n_{i}>}^{2})-\frac{U}{2}\sum_{i}n_{i}-\mu\sum_{i}n_{i}.
\end{align}

We simulate the effective model Hamiltonian $(H_{eff})$ using semi-classical
Monte Carlo (s-MC)~\cite{Mukherjee,Chakraborty1, Mukherjee1,Patel,Tiwari} method
by diagonalizing the fermionic sector in the background of fix $\{{\bf{m}}_{i}\}$
and $\{<n_{i}>\}$ configurations. The classical $\{{\bf{m}}_{i}\}$ variables
are updated by visiting every lattice site sequentially by implementing metropolis
algorithm. We determine $\{<n_{i}>\}$ self-consistently at every $10th$ step of
the MC system sweep which is then used in the next $10$ MC sweeps. We measure
physical quantities at every $10th$ step after the thermal equilibrium to
discard illicit correlation in the data. We compute the effective Hamiltonian
in a large system size $8\times8\times12$  with the help of travelling
cluster approximation (TCA)~\cite{Kumar,Pradhan,Chakraborty,Halder}
based Monte Carlo technique using $4\times4\times8$ TCA cluster.

\begin{figure}[!t]
\centerline{
\includegraphics[width=8.5cm,height=6.30cm,clip=true]{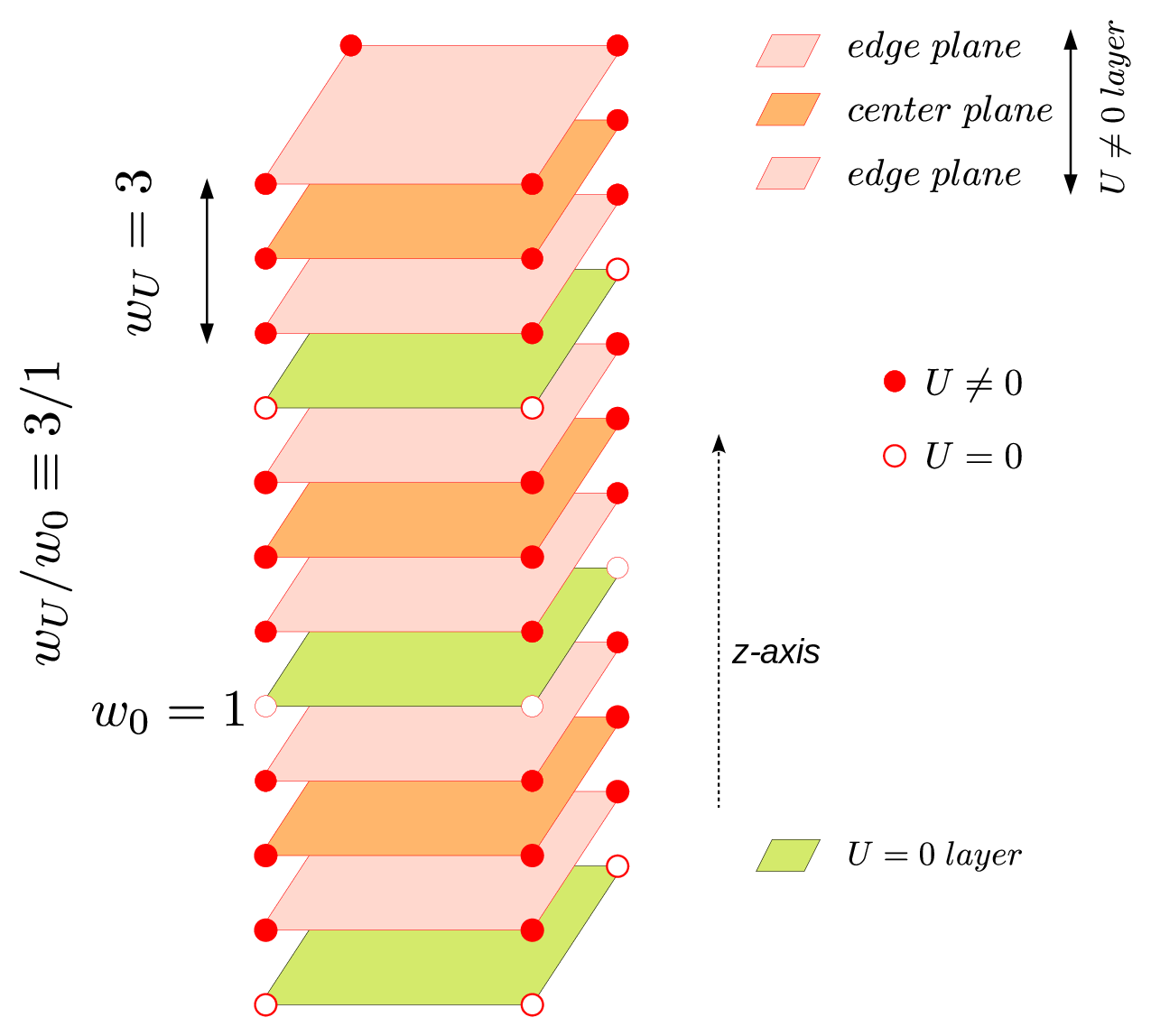}}
\caption{
Schematic view of the superlattice system. Red filled (open) circles
represent the correlated (uncorrelated) sites with $U \neq 0$ ($U = 0$). The
width of the correlated (uncorrelated) layer is $w_{U}$ ($w_{0}$) i.e. it
contains $w_{U}$ ($w_{0}$) number of planes. These correlated and uncorrelated
layers are periodically arranged along the z-axis and denoted as $w_{U}/w_{0}$
SL. The edge and center planes of the correlated layer are represented by
different colors. Total width of the superlattice along the z-axis is fixed;
$L_{z} = 12$. In this schematic of 3/1 SL $w_{U}/w_{0} \equiv 3/1$ and
$3\times(w_{U}+w_{0}) = 12$.
} 
\label{sch_sl}
\end{figure}

In the bulk system (where all the sites have finite $U$), the ground state
remains in an antiferromagnetic insulating (AF-I) state for $U > 0$ and in a
paramagnetic metallic (PM-M) state for $U = 0$, at half-filling. We have
designed AF/PM (correlated/uncorrelated) superlattices (SLs) where the
AF layer ($U \ne 0$) of width $w_{U}$ and the PM layer ($U = 0$)
of width $w_{0}$ are periodically arranged to form the superlattices.
We call the superlattices as $w_{U}/w_{0}$ SLs, illustrated schematically
in Fig.~\ref{sch_sl}. We define $x$ ($= \frac{w_{U}}{w_{U}+w_{0}}$)
as the fraction of correlated planes in the $w_{U}/w_{0}$ SL.

\section{Physical Observables}
We evaluate various physical observables to investigate the magnetotransport
properties of the whole antiferromagnetic/paramagnetic superlattices. For all
superlattices we calculate the layer resolved (separately for correlated layers
and uncorrelated layers) observables. In some cases we also calculate plane
resolved properties of a given layer (correlated or uncorrelated).

The system averaged magnetization squared is calculated as
$M =\langle (n_{\uparrow}-n_{\downarrow})^{2}\rangle =\langle n \rangle - 2\langle n_{\uparrow}n_{\downarrow}\rangle$
where $\langle n \rangle = \langle n_{\uparrow}+n_{\downarrow}\rangle$. The
average magnetic moment irrespective of its direction is inferred from this
indicator $M$. The angular brackets represent quantum mechanical and 
thermal averages throughout the Monte Carlo generated equilibrium configurations.
To analyze the long range antiferromagnetic order in superlattices we calculate
the structure factor S({\bf{q}}) for ${\bf{q}}=(\pi, \pi, \pi)$ as follows:
\[S({\bf{q}})=\frac{1}{N^{2}}\sum_{i,j}\langle{\bf{S}}_{i}.{\bf{S}}_{j}\rangle
e^{i{\bf{q}}.({\bf{r}}_{i}-{\bf{r}}_{j})}\] where $N$ is the total number of
sites of the system and $i$ and $j$ run all over the sites of the system. 
We also calculate the specific heat of the system by differentiating the total 
energy with respect to temperature, $C_{v}(U,T)=\frac{dE(U,T)}{dT}$. We apply
central difference formula to evaluate the specific heat numerically.

Density of states (DOS) of the system at frequency $\omega$ is determined by using
the expression $D(\omega)=\sum_{\alpha}\delta(\omega-\epsilon_{\alpha})$, where $\epsilon_{\alpha}$ 
are the single particle eigen values and $\alpha$ runs over the total number ($=2N$) of 
eigen values of the system. In our simulation, Lorenzian representation of the delta
function with broadening $\sim BW/2N$ ($BW$ is the bare bandwidth) is
applied to evaluate the DOS. 

The longitudinal (along $z$-axis) and transverse (along $x$-axis) resistivity of the
superlattice are calculated by using dc limit of the optical conductivity through 
Kubo-Greenwood formalism~\cite{Pradhan,Mahan,Kumar1}, represented by

\[\sigma(\omega)=\frac{A}{N}\sum_{\alpha, \beta}(n_{\alpha}-n_{\beta})
\frac{{|f_{\alpha \beta}|}^{2}}{\epsilon_{\beta}-\epsilon_{\alpha}}
\delta[\omega-(\epsilon_{\beta}-\epsilon_{\alpha})]\]

\noindent
where $A=\pi e^{2}/\hbar a$ ($a$ is the lattice parameter). $f_{\alpha \beta}$
represents the matrix elements of the current operator
${\hat{j}}_{z}=it\sum_{i,\sigma}
(c_{i,\sigma}^{\dagger}c_{i+z,\sigma}-c_{i+z,\sigma}^{\dagger}c_{i,\sigma})$
or, ${\hat{j}}_{x}=i\sum_{i,\sigma}(c_{i,\sigma}^{\dagger}c_{i+x,\sigma}-
c_{i+x,\sigma}^{\dagger}c_{i,\sigma})$
between the eigen states $|\psi_{\alpha}>$ and $|\psi_{\beta}>$ with corresponding
eigen energies $\epsilon_{\alpha}$ and $\epsilon_{\beta}$, respectively
and $n_{\alpha} = \theta(\mu - \epsilon_{\alpha})$ is the Fermi function
associated with the single particle energy level $\epsilon_{\alpha}$. Afterwards the 
dc conductivity is computed by averaging over the low-frequency interval as follows:
\[ \sigma_{av}(\Delta\omega)=\frac{1}{\Delta\omega}\int_{0}^{\Delta\omega}\sigma(\omega) d\omega\]
where $\Delta\omega$ is selected three to five times larger than the average eigen value 
separation of the system, defined as the ratio of the bare bandwidth to the total number
of eigen values. All the physical parameters like $U$, $T$, $\omega$ are measured
in units of $t$.

%***************************************************************************
%***************************************************************************

%***************************************************************************
\begin{figure}[!t]
\centerline{
\includegraphics[width=8.5cm,height=6.3cm,clip=true]{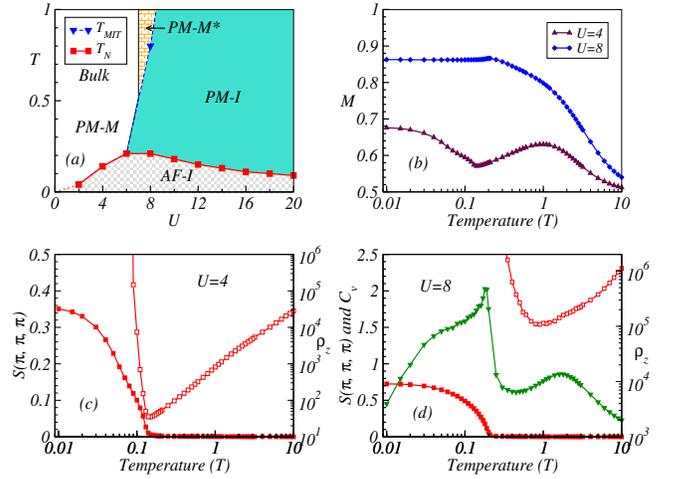}}
\caption{
(a) $U-T$ phase diagram of the bulk system ($U \ne 0$ for all the sites).
The system directly converts to an AF-I state from a PM-M state for $U < 8$.
A PM-I phase intervenes between the PM-M* and the AF-I phases for $U \ge 8$.
For details about the characterization of these phases please see the text.
(b) The local magnetic moment $M$ (for $U = 8$) increases with decreasing
temperature and saturates at low temperature, displaying a small peak near
$T_{N}$. For $U = 4$, $M_{U}$ decreases from $T \sim 1$ down to $T \sim T_{N}$
and then starts to increase below $T_{N}$.
(c) Temperature dependence of $S(\pi, \pi, \pi)$ (left axis) and $\rho_{z}$
(right axis) for $U = 4$ show that the antiferromagnetic transition and the
metal-insulator transition occur simultaneously at the same temperature
(i.e. $T_{N}$ = $T_{MIT}$).
(d) For $U = 8$ the $T_{MIT}$ is larger than $T_{N}$ and as a result PM-I
phase appears between the PM-M* and the AF-I phases as shown in (a). The high-T
$C_{v}$ peak is consistent with the metal-insulator transition, while the
low-T $C_{v}$ peak corresponds to the antiferromagnetic transition.
}
\label{pd_bulk}
\end{figure}
%***************************************************************************

%***************************************************************************
\begin{figure}[!t]
\centerline{
\includegraphics[width=8.5cm,height=6.3cm,clip=true]{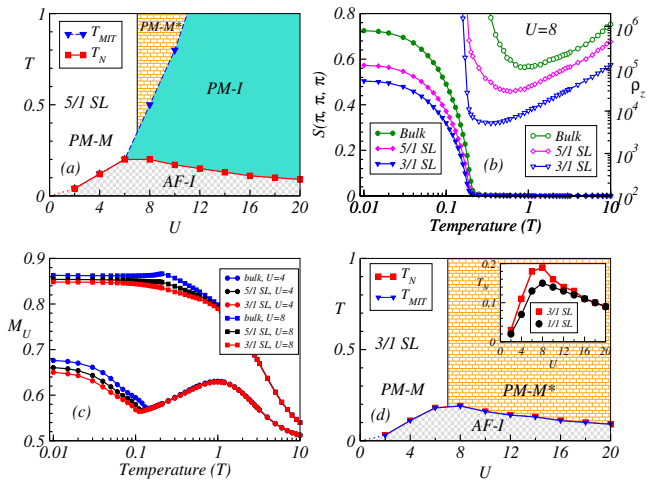}}
\caption{
(a) The $U-T$ phase diagram of the 5/1 superlattice: Area of the PM-I phase
shrinks whereas the metallic phase (PM-M*) gets enlarged as compared to the
bulk system phase diagram.
(b) Temperature dependence of $S(\pi, \pi, \pi)$ (left axis) and $\rho_{z}$
(right axis) are plotted for bulk ($x = 1$),  5/1 SL ($x \sim 0.83$) and
3/1 ($x = 0.75$) SL for $U = 8$. The $T_{MIT}$ decreases as we move from bulk
to 5/1 to 3/1 SL whereas the $T_{N}$s remain more or less same.
The $T_{MIT}$ matches with the $T_{N}$ only for 3/1 SL.
(c) Temperature evolution of the local moments in the correlated layers
($M_{U}$) are plotted for bulk, 5/1 and 3/1 SLs using two different $U$ values
($U$ = 4 and 8). Low temperature saturation magnetic moment increases as one
moves from 3/1 SL to 5/1 SL to bulk systems. For $U = 4$ a dip near the $T_{N}$
is present in all three systems.
(d) The PM-I phase disappears in $U-T$ phase diagram for 3/1 SL. The PM-I
phase is covered-up by PM-M* phase. The 3/1 SL directly transforms from PM-M
(or PM-M*) state to AF-I state. For more details please see the texts.
}
\label{pd_sl}
\end{figure}
%***************************************************************************

\section{$U-T$ Phase diagrams for various superlattices }

First we briefly discuss the physics of the bulk system ($U \neq 0$ for all the
sites) before analyzing the intriguing phenomena in AF/PM superlattices.
We present the $U-T$ phase diagram for bulk system in Fig.~\ref{pd_bulk}(a). The
ground state of the bulk system shows a G-type antiferromagnetic (G-AF) insulating
phase at low temperatures for all $U$ values. Please notice that the Neel
temperature ($T_{N}$) varies nonmonotonically with $U$ values. Specifically, $T_{N}$
increases with $U$ up to $U = 8$ and decreases thereafter. For $U < 8$ the system
transits from a paramagnetic metallic (PM-M) phase to an AF insulating phase
(AF-I) with decreasing the temperature and the metal insulator transition
temperature ($T_{MIT}$) coincides with the $T_N$. On the other hand, for $U = 8$
or more the system encounters a paramagnetic insulating (PM-I) region
above the $T_{N}$. As we increase the temperature further the system crosses over
to a slightly different kind of paramagnetic metallic phase (abbreviated as PM-M*)
above $T_{MIT}$ (denoted by dashed line). In order to characterize the properties
of PM-M (for $U$ $\le$ 6) and PM-M* (for $U$ $\ge$ 8) phases we focus on the
$M$ vs $T$ plot (see Fig.~\ref{pd_bulk}(d)).  In PM-M* phase (for $U = 8$ case),
the magnetic moment $M$ increases with decreasing temperature whereas
the magnetic moment $M$ decreases in the PM-M phase (for $U = 4$ case) as we
go below $T = 1$ down to $T_N$ in our phase diagram. In other words local magnetic
moments are preformed in PM-M* phase at high temperatures ($\sim$ 1) unlike the
PM-M phase. Although not necessary, we differentiate PM-M and PM-M* phases in
the phase diagram for brevity, which will be useful in discussing the phase
diagrams of SL systems.

We plot the magnetic and transport properties in Fig.~\ref{pd_bulk}(b)-(d) that
we used to set up the $U-T$ phase diagram presented in Fig.~\ref{pd_bulk}(a).
Mainly, we focus on two $U$ values ($U = 4$ and $8$) which represent two different
regimes in our $U-T$ phase diagram. As we mentioned earlier the structure factor
calculations show that the $T_{N}$ for $U = 4$ is smaller than the $T_{N}$ for
$U = 8$ [see Figs.~\ref{pd_bulk}(c) and (d)]. The $T_{N}$ and the $T_{MIT}$ of
the bulk system are equal to each other for $U = 4$. On the other hand the
$T_{MIT}$ is much larger than $T_{N}$ for $U = 8$. A paramagnetic insulating
(PM-I) phase intervenes between the PM-M* and AF-I phase at larger $U$ values.
We also plot the specific heat $C_{v}$ vs temperature for $U = 8$ case in
Fig.~\ref{pd_bulk}(d). The high-temperature $C_{v}$ peak associated with
charge-fluctuations coincides with the $T_{MIT}$ whereas the low-T peak of $C_{v}$ 
associated with spin-fluctuations matches well with the $T_{N}$~\cite{Duffy,Paiva}. 
It is also apparent that, for $U = 4$, the local magnetic moment
$M$ in PM-M phase decreases upon decreasing the temperature from $T \sim 1$ up to
$T_{N}$ with the enhancement of the metallicity as shown in Fig.~\ref{pd_bulk}(b).
Below $T_{N}$ the local magnetic moment $M$ increases again with decreasing the
temperature. For $U = 8$ the magnetic moment $M$ gradually increases with decreasing
the temperature barring a small peak around $T_N$ and saturates at low temperatures.
The small drop just below $T_N$ is due to the delocalization of the electrons assisted
by virtual hopping facilitated by the antiferromagnetic correlations. The local moment
for both $U = 4$ and $U = 8$ converge to the asymptotic value $M = 0.5$ at very high
temperatures. So, the system directly transforms to an AF-I state from PM-M 
state upon cooling for $U < 8$ whereas the system switches to an AF-I from PM-M*
via a PM-I state for larger $U$ values as shown in our phase diagram. Overall, the bulk
phase diagram obtained in our semi-classical Monte-Carlo (s-MC) approach is consistent with
the DQMC~\cite{Blankenbecler,Rohringer}, 2D cluster-DMFT (C-DMFT)~\cite{Fratino},
3D cluster-DMFT (C-DMFT)~\cite{Sato} with vortex corrections and QMC
simulations~\cite{Staudt}.

Next, we discuss the $U-T$ phase diagram for different AF/PM superlattices, namely
5/1, 3/1 and 1/1 SLs where $x \ge 0.5$ ($x$ denotes fraction of correlated planes).
We start our analysis with the  5/1 SL [see Fig.~\ref{pd_sl}(a)]. The nonmonotonic
behavior of $T_{N}$ with $U$ remains intact for 5/1 SL, similar to the bulk systems
although the $T_{N}$ values decreases very slightly. The $T_{MIT}$ for $U < 8$
also coincides with the $T_{N}$ (i.e. the SL system is metallic above the $T_{N}$) and
the local magnetic moment decreases as we approach the $T_{N}$ from the high
temperature [see Fig.~\ref{pd_sl}(c) for $U = 4$ case]. As a result the PM-M phase,
observed in 5/1 SL, remains very similar to the bulk system. PM-M* phase is enlarged
for intermediate $U$ values as compared to the bulk phase diagram
[please compare Fig.~\ref{pd_sl}(a) and Fig.~\ref{pd_bulk}(a)].
To analyze the enlarged PM-M* phase further we compare the $T_N$ and $T_{MIT}$
obtained from the antiferromagnetic correlations and the resistivity curves,
respectively in Fig.~\ref{pd_sl}(b). Interestingly for $U = 8$ reduction in $T_{N}$
is hardly noticeable. On the other hand the $T_{MIT}$ decreases considerably
due to the insertion of the uncorrelated layers. But, $T_{MIT}$ still remains
larger than the $T_N$. Magnetotransport properties of individual layers
(comprised of $U = 8$ and $U = 0$) will be discussed later. In this PM-M*
phase the magnetic moments in the correlated layer $M_{U}$ gets
more and more localized as we decrease the temperature
(see $U = 8$ case in Fig.~\ref{pd_sl}(c)] unlike the $U = 4$ case.

Now we outline the magnetotransport properties and the phase diagram of 3/1 SL. 
The $T_N$ profile of 3/1 SL remains more or less the same to that of 5/1 SL
(see Fig.~\ref{pd_sl}(d)). We show the comparison of the $T_N$ of 3/1 and 5/1
SLs in Fig.~\ref{pd_sl}(b) for $U = 8$. Interestingly, the $T_{MIT}$ coincides
with the $T_{N}$ at $U = 8$ in 3/1 SL (unlike 5/1 SL and the bulk systems) as
shown in Fig.~\ref{pd_sl}(b). In fact the $T_{N}$ also coincides with $T_{MIT}$
at larger $U$ values. As a result the intervening PM-I phase seen between PM-M*
and AF-I in 5/1 SL is overtaken by PM-M* phase. So, one directly enters into an
antiferromagnetic insulating (AF-I) phase from a paramagnetic metallic (PM-M*)
phase in the 3/1 SL system with decrease in temperature. Similar phase diagram
is also obtained for the 1/1 SL but the $T_{N}$ [see the inset of Fig.~\ref{pd_sl}(d)]
gets reduced as compared to the 3/1 SL.

\section{Long range antiferromagnetic ordering: Mutual cooperation between
correlated and uncorrelated layers}

\begin{figure}[!t]
\centerline{
\includegraphics[width=8.5cm,height=6.30cm,clip=true]{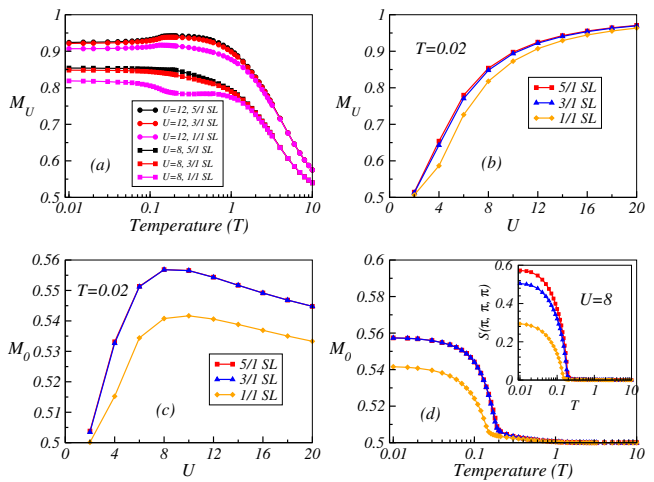}}
\caption{
(a) Temperature evolution of the local moments in the correlated layer ($M_{U}$) 
of three different superlattices (5/1, 3/1 and 1/1 SLs) are plotted for $U = 8$
and $U = 12$. The local moment increases monotonically with decreasing temperature
and saturates at low temperature. For both values of $U$, the local moment in
the 1/1 SL is smaller than the 3/1 and 5/1 SLs. (b) At low temperatures ($T = 0.02$),
the local moment in the correlated layer ($M_{U}$) increases with $U$ and
eventually saturates for larger $U$ values. The local moment $M_{U}$ in 1/1 SL
is smaller than the 3/1 and 5/1 SLs for intermediate $U$
values. (c) The induced moment in the uncorrelated layer ($M_{0}$) for 5/1,
3/1 and 1/1 SLs exhibits a nonmonotonic behavior with $U$ at low temperatures
($T = 0.02$). The induced moment in the 1/1 SL is considerably smaller than the
3/1 and 5/1 SLs. (d) For $U = 8$ the onset temperature of the induced moment
in the uncorrelated layer (main panel) is consistent with the antiferromagnetic
ordering temperature of the SL systems (inset).
} 
\label{M_UT}
\end{figure}

In this section we analyze the magnetic and transport properties of the
individual layers (i.e. separately for correlated ($U \neq 0$) and uncorrelated
($U = 0$) layers) of our SL systems to establish a mutual cooperation between
the magnetic ordering of the correlated and uncorrelated layers. We plot the
local moment $M_{U}$ profile of the correlated layers for $U = 8$ and $U = 12$
for 5/1, 3/1 and 1/1 SLs in Fig.~\ref{M_UT}(a). The local moment $M_U$
(calculated exclusively for correlated layers) gradually increases with
decreasing the temperature for both the $U$ values and saturates at low
temperatures. The local moment in the correlated layer for 1/1 SL is visibly
smaller than the 5/1 and 3/1 SLs below $T = 1$. The bidirectional coupling
of the correlated plane with the adjacent uncorrelated planes in 1/1 SL suppresses
the local moment in the correlated layers. As we increase the $U$ values the
local moment in the correlated layer for 1/1 SL approaches to that of
the 3/1 and 5/1 SLs due to enhancement of the localization in the correlated
layers. At low temperature $T = 0.02$, the local moment of the correlated layers
increases monotonically with increase of $U$ (since the double occupancy reduces 
with increase of $U$) and saturates at large $U$ values as shown in Fig.~\ref{M_UT}(b).
But, the local moment $M_{U}$ in the correlated layer for 1/1 SL is smaller than
the 3/1 and 5/1 SLs for intermediate $U$ values as mentioned above.
Otherwise, the qualitative nature of the local moments profile in the
correlated layers remains similar for the 5/1, 3/1 and 1/1 SLs.

%***************************************************************************
%***************************************************************************

On the other hand, the average induced moment in the uncorrelated layer ($M_{0}$)
shows nonmonotonic behavior with increase of $U$ values and the maximum moment
is obtained for $U = 8$ for all the three SLs [see Fig.~\ref{M_UT}(c)]. These
induced moments in the uncorrelated layer are generated due to the kinetic hopping
of the charge carriers from the correlated layers. In addition, it is apparently
clear that the induced moment $M_{0}$ in uncorrelated layers for 1/1 SL is
substantially smaller than the 3/1 and 5/1 superlattices.
The induced magnetic moment in the uncorrelated layer is similar to the
experimentally observed induced magnetic moment in the paramagnetic $Pd$ layer
in NiO/Pd multilayers~\cite{Saha,Manago,Manago1,Manago2,Manago3} and
the induced magnetic moment in $Cu$ layer in CuO/Cu multilayers~\cite{Munakata}.

In order to extract the temperature scale at which magnetic moments are actually
induced in the uncorrelated layers we plot the local moment of uncorrelated layers
vs temperature for 5/1, 3/1, and 1/1 SLs for $U = 8$ case in Fig.~\ref{M_UT}(d).
Local moments are induced in the uncorrelated layer at very low temperatures
(the temperature where $M_0$ becomes greater than 0.5) as compared to the
correlated layers. The onset temperature of the induced moment in the uncorrelated
layer ($M_0$) around $T = 0.2$ matches well with the $T_N$ of these SL systems
(see the inset of Fig.~\ref{M_UT}(d)). So our calculations show a one-to-one
correspondence between antiferromagnetic ordering and onset of induced moments
in uncorrelated layers in the SL systems.

\begin{figure}[!t]
\centerline{
\includegraphics[width=8.5cm,height=6.30cm,clip=true]{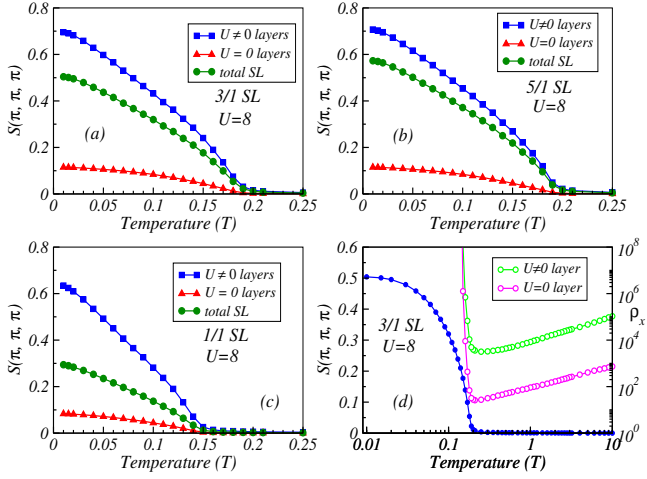}}
\caption{
Comparison of the antiferromagnetic ordering temperature (from $S(\pi, \pi, \pi)$
vs $T$ plots) between the correlated layers ($U \neq 0$), uncorrelated layers
($U = 0$), and total SL (including both correlated and uncorrelated layers)
for $U = 8$ in (a) 3/1 SL,  (b) 5/1 SL, and (c) 1/1 SL.
Individual correlated and uncorrelated layers exhibit antiferromagnetic
transitions simultaneously which matches with $T_N$ of whole SL system.
A mutual cooperation between correlated and uncorrelated layers establishes
the long range AF order in the whole superlattices. (
d) In-plane resistivity $\rho_{x}$ of the correlated and uncorrelated layers
for 3/1 SL at $U = 8$ show that the MIT occurs at $T_{N}$. For a better
clarity we have also re-plotted temperature dependence of $S(\pi, \pi, \pi)$
for 3/1 SL in the same figure.
} 
\label{afm_layer}
\end{figure}

To further analyze the correspondence between the onset of antiferromagnetic
ordering in correlated and uncorrelated layers we plot $S(\pi, \pi, \pi)$
structure factor of both the layers separately and compare it with the
$S(\pi, \pi, \pi)$ structure factor of the total SL system in
Figs.~\ref{afm_layer}(a)-(c). The structure factors show that the antiferromagnetic
order appears at the same temperature for both correlated layers ($U \ne 0$ layers)
and uncorrelated layers ($U = 0$ layers) and matches well with the $T_N$ of
the whole superlattice system. This also indicates that the correlations among
the induced moments in the uncorrelated layer play an important role in
mediating the long range interactions between the correlated layers. So, overall
the cooperation between the correlated and uncorrelated layers helps
to sustain the global long range AF order in the superlattices. We also plot
the layer resolved resistivities for the 3/1 SL in Fig.~\ref{afm_layer} (d). 
$S(\pi, \pi, \pi)$ of whole 3/1 SL is re-plotted in the same figure for
comparing the transition temperatures. The $T_{MIT}$ obtained from correlated
layers and uncorrelated layers are equal to each other and coincides with the
$T_N$ for the SL system.

%***************************************************************************
%***************************************************************************

\section{Plane resolved transport properties of the superlattices}

\begin{figure}[!t]
\centerline{
\includegraphics[width=8.5cm,height=6.30cm,clip=true]{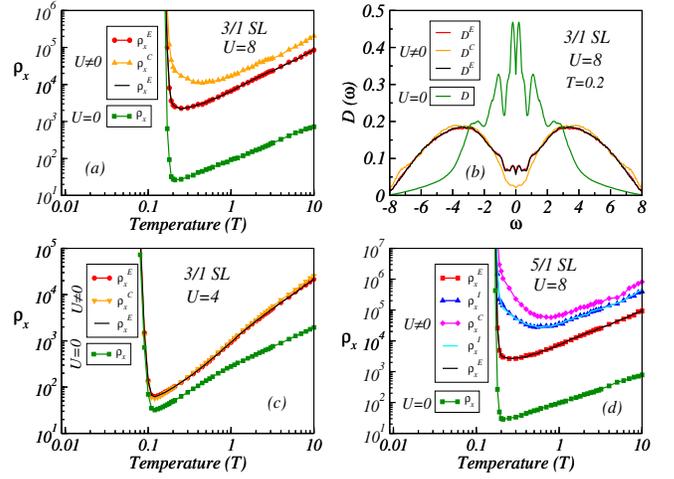}}
\caption{
In-plane resistivity ($\rho_{x}$) vs temperature ($T$) for all the constituent
planes (three correlated planes and one uncorrelated plane) of the 3/1 SL are
plotted for (a) $U = 8$ and (c) $U = 4$. For $U = 8$, both the $T_{MIT}$ as well
as the magnitude of the resistivity ($\rho_{x}$) of the edge correlated planes
are smaller than that of the center correlated plane (i.e. $T_{MIT}^{E} < T_{MIT}^{C}$
and $\rho_{x}^{E} < \rho_{x}^{C}$). The $T_{MIT}$ of the edge correlated
plane matches with $T_{N}$. In case of $U = 4$ all the individual planes show
MIT exactly at $T_{N}$. In fact, for $U =4$, the resistivity curves obtained
from the the correlated planes are overlapping on top of each other. The
resistivity of the uncorrelated plane ($U = 0$) is much smaller than the
correlated planes for both $U$ values.
(b) Plane resolved density of states (DOS) of the 3/1 SL at $T = 0.2$ (around
$T_N$): The weight of the DOS at the Fermi level ($\omega = 0$) for the edge
correlated planes are larger than the central correlated plane ($D^{E} > D^{C}$).
But, the weight of the DOS of the uncorrelated plane ($U = 0$) at the Fermi level
is very predominant among the four planes. Our plane resolved DOS calculations
correspond well with the in-plane resistivity calculations shown in (a).
(d) In-plane resistivity of the individual planes in 5/1 SL at $U = 8$ shows
similar characteristics behavior to that of 3/1 SL in (a). Both the $T_{MIT}$
and magnitude of $\rho_{x}$ increase as we move from the edge plane to the
center plane inside the correlated layer. Also the $\rho_{x}$ in the uncorrelated
plane ($U = 0$) is much smaller than the correlated planes. Here E, I, and C
stand for edge, intermediate and center planes, respectively.
} 
\label{rho_31}
\end{figure}

Here, we investigate the plane resolved transport properties of
the correlated and uncorrelated layers of the superlattices.
First we focus on 3/1 SL where only one uncorrelated plane ($U = 0$) is
intercalated between the correlated layers made up of three correlated planes.
For this we evaluate the in-plane resistivity ($\rho_{x}$) of individual planes
(three correlated planes and one uncorrelated plane) for $U=8$ as shown in
Fig.~\ref{rho_31}(a). Interestingly, the metal-insulator transition temperature of
the edge correlated planes ($T_{MIT}^{E}$) is smaller than the center
($T_{MIT}^{C}$) correlated planes (i.e, $T_{MIT}^{E}<T_{MIT}^{C}$) in 3/1 SL.
It is clearly observed that the center correlated plane exhibits MIT above $T_N$
whereas the $T_{MIT}$ of both the edge planes coincides with the $T_N$. The $T_{MIT}$
of the uncorrelated plane also matches with the $T_N$. Expectedly the value
of $\rho_{x}$ in the uncorrelated plane is much smaller than the correlated plane.

%***************************************************************************
%***************************************************************************

%***************************************************************************
\begin{figure}[!t]
\centerline{
\includegraphics[width=8.5cm,height=4.0cm,clip=true]{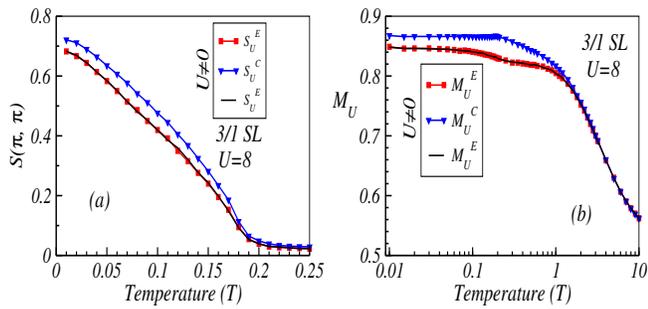}}
\caption{
(a) Temperature dependence of $S(\pi, \pi)$ of the individual correlated planes
($U \neq 0$) of 3/1 SL at $U = 8$ are plotted to compare the $T_N$s.
Antiferromagnetic transition temperatures remain more or less same for the
individual constituent planes (two edge planes and one center plane).
$S(\pi, \pi)$ of the edge plane is smaller than the center plane at low
temperatures.
(b) The local moments in the edge correlated planes of the 3/1 SL are smaller
than the center correlated plane ($M_{U}^{E} < M_{U}^{C}$) at low temperatures
corresponds well with the structure factor calculations presented in (a).
} 
\label{afm_plane}
\end{figure}
%***************************************************************************

Next, we calculate plane-resolved density of states (DOS) to further
deliberate the transport properties. Plane-resolved DOS of the correlated layer at
$T = 0.2$ (see Fig.~\ref{rho_31}(b)) indicate that weight of the DOS of the
edge correlated planes at the Fermi level (set at $\omega = \varepsilon_{F} = 0$)
are larger as compared to the central correlated plane (i.e. $D^{E} > D^{C}$
at $\omega = 0$). Expectedly the weight of DOS of the uncorrelated plane
at Fermi level is significantly greater than all the correlated planes.
Hence, our plane-resolved DOS calculations are coherent with the in-plane
resistivity calculations of the individual planes of the superlattice.

For smaller $U$ values all the planes (comprised of three correlated planes and
one uncorrelated plane) show metal-insulator transition at the same temperature
[see Fig.~\ref{rho_31}(c)] and this temperature coincides with the
$T_N$. In fact, the resistivity curves of all three correlated planes overlap
with each other. This shows that all the correlated planes are equally affected
by the insertion of uncorrelated layer for smaller $U$ values where moments are
much more delocalized as compared to higher $U$ values.
Our magnetotransport results qualitatively follow the QMC~\cite{Jiang} and
DQMC~\cite{Euverte} studies of the correlated superlattices where it was shown that 
the correlated layer is affected by the uncorrelated layer and the effect
increases with decrease of $U$. But, the detailed transport properties were
not reported earlier. The transport properties of 5/1 SL [see Fig.~\ref{rho_31}(d)]
remains qualitatively similar to that of 3/1 SL where the $T_{MIT}$ increases
as we move from the edge plane to the center plane.

In our 3/1 SL the correlated layer is made up of three planes. So the obvious
question arises at this point: Does all the three correlated planes (two edge
planes and one center plane) that constitute the correlated layer align
antiferromagntically at the same temperature or not? To answer this question
we plot the antiferromagnetic structure factor of all the three correlated
planes separately using $U = 8$ in Fig.~\ref{afm_plane}(a). The AF correlations
of individual planes vanish at the same temperature. But, the reduction
in the low-temperature saturation value in edge plane as compared to center
plane is very clear. This may be due to the larger magnetic moment in the center
plane than the edge planes. To confirm this we plot the local magnetic moment
$M_U$ of individual correlated planes vs temperature in Fig.~\ref{afm_plane}(b).
In fact, the $M_U$ for the center plane is larger than the edge plane at low
temperatures.

The uncorrelated plane affects the local magnetic moments $M_U$ of the correlated
planes differently ($M_U$ is larger for central plane) due to the proximity effect
as we discussed earlier. Apparently the $M_U$ profiles indicate that the resulting
effective $U$ ($U_{eff}$) of center plane is larger than the edge plane. This is also
supported from resistivity plots where $T_{MIT}$ of the center plane is larger than
the edge plane (see Fig.~\ref{rho_31} (a)).
From all these analysis one would naively expect that the
$T_N$ of the edge plane should be smaller than the center plane but this difference
is beyond the resolution of our calculations and we get more or less same $T_N$
[see Fig.~\ref{afm_plane}(a)].

%***************************************************************************
\begin{figure}[!t]
\centerline{
\includegraphics[width=8.5cm,height=6.30cm,clip=true]{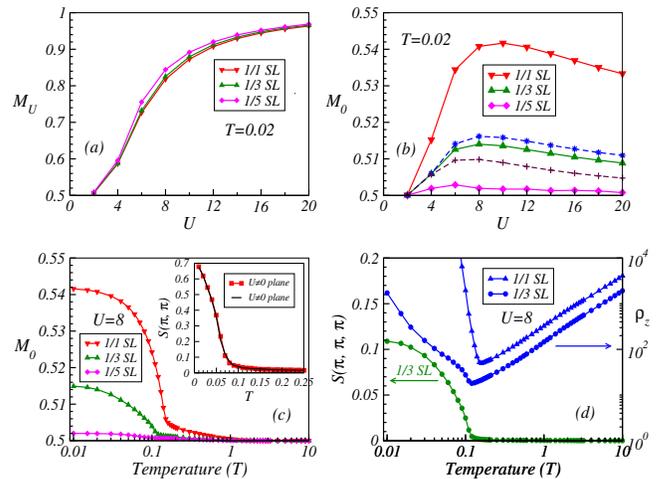}}
\caption{Magnetotransport properties of 1/3, and 1/5 SLs. We re-plot the
corresponding data of 1/1 SL for comparison. (a) At low temperatures ($T = 0.02$)
the local moment in the correlated layer increases monotonically with $U$ and
saturates at higher values of $U$ for both 1/3 and 1/5 SLs similar to 1/1 SL.
(b) The induced  moment in the uncorrelated layer at $T = 0.02$ shows
nonmonotonic behavior with $U$ for 1/3 SL. However, for the 1/5 SL, the induced
moment in the uncorrelated layer is negligibly small for all values of $U$.
Induced moment profile of the edge (center) uncorrelated plane for 1/3 SL
is also plotted by blue (maroon) dashed line. Expectedly, induced moment
in the edge plane is larger than the center plane.
(c) Temperature dependence of $M_{0}$ for $U = 8$ show that there is a clear
onset temperature of the induced moment in the uncorrelated layer in 1/3 SLs,
similar to 1/1 SL. There is no clear onset temperature for 1/5 SL. In the inset,
temperature dependence of the structure factor $S(\pi, \pi)$ for the correlated
planes separated by five uncorrelated planes in 1/5 SL are plotted for $U = 8$.
Individual correlated planes show antiferromagnetic transitions, but the
long range antiferromagnetic order between them is not feasible due to the
insufficient induced moment in the uncorrelated layer, which acts as a mediator.
(d) Temperature dependence of $S(\pi, \pi, \pi)$ for 1/3 SL at $U = 8$ (left axis)
shows a clear antiferromagnetic transition. The $T_{MIT}$ obtained from
$\rho_{z}$ vs $T$ plot matches well with the $T_{N}$ for 1/3 SL. The value of
$\rho_{z}$ in 1/3 SL is smaller than the 1/1 SL due to the insertion of a
thicker uncorrelated layer between the correlated layers.
} 
\label{rho_13}
\end{figure}
%***************************************************************************

%***************************************************************************
\begin{figure}[!t]
\centerline{
\includegraphics[width=8.5cm,height=6.30cm,clip=true]{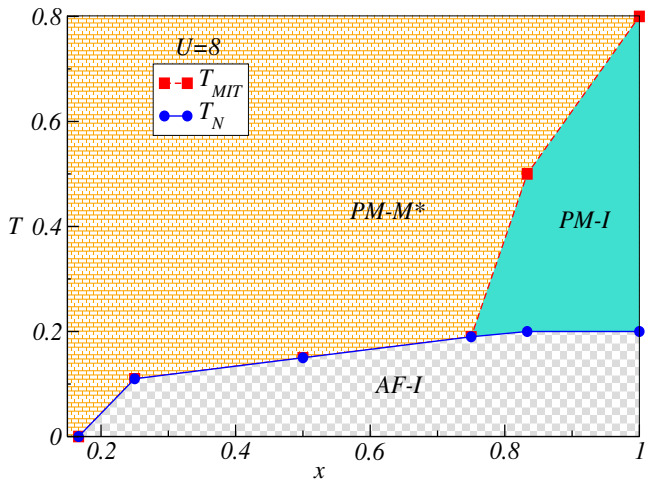}}
\caption{
$x-T$ phase diagram comprises different superlattices (where $w_U$ = 1 or/and
$w_0$ = 1) for $U = 8$.  Here, $x = \frac{w_{U}}{w_{U}+w_{0}}$ is the fraction
of correlated planes ($U \neq 0$). $x = 1$ represents the bulk system where all
the sites have $U \neq 0$. We use the results from 1/5 SL ($x \sim 0.17$), 1/3
SL ($x \sim 0.25$), 1/1 SL ($x \sim 0.5$), 3/1 SL ($x \sim 0.75$), 5/1 SL
($x \sim 0.83$) and bulk ($x \sim 1$) to plot the phase diagram. Both $T_{N}$
(blue line) and $T_{MIT}$ (red dashed line) decrease as we decrease the $x$
value. System encounters a PM-I phase between the AF-I and $PM$-$M^{*}$ phases
for $x > 0.75$. The $T_{MIT}$ merges with the $T_{N}$ for $x \leq 0.75$ i.e.
the system directly transits from PM-M* to AF-I state upon cooling.  At
$x \sim 0.17$ (for 1/5 SL) the long range AF order collapses due to the
insufficient induced moments in the uncorrelated layers as discussed
earlier in Fig.~\ref{rho_13}(c).
} 
\label{pd_u8}
\end{figure}
%***************************************************************************

\section{Magnetotransport properties of 1/3 and 1/5 superlattices}

Now we briefly analyze the transport and magnetic properties of 1/3 and 1/5 SLs
to present a $x-T$ diagram. We plot the local moment of the correlated layers for
1/1, 1/3 and 1/5 SLs with different $U$ values at $T = 0.02$. The local moments of
the correlated layers $M_U$ increases monotonically with increase of $U$ and
saturates at large $U$ values as shown in Fig.~\ref{rho_13}(a) for all three SLs.
So, the qualitative nature of the local moments profile in the correlated planes
remains the same as we increase the thickness of uncorrelated layer. We also
plot the induced magnetic moment in uncorrelated layers in Fig.~\ref{rho_13}(b).
Induced magnetic moment shows nonmonotonic behavior for 1/1 and 1/3 SLs.
But, induced magnetic moment decreases drastically for 1/5 SL.
The average induced moment $M_0$ in the edge and center uncorrelated ($U = 0$)
planes are also plotted (see Fig.~\ref{rho_13}(b)) for 1/3 SL. The nonmonotonic
behavior remains intact for both edge and center plane and expectedly the
induced moment in the edge uncorrelated plane is larger than the center
uncorrelated plane.

The induced moments in the uncorrelated layer decreases as we
shift from 1/1 to 1/3 SL as shown in Fig.~\ref{rho_13}(b). But, there is a clear
onset temperature of the induced moment in the uncorrelated layer ($M_0$) around
$T = 0.1$ for 1/3 SL as shown in Fig.~\ref{rho_13}(c). So, one expect long range
antiferromagnetic correlations in 1/3 SL. In fact, magnetic structure factor
S($\pi$,$\pi$,$\pi$) for 1/3 SL (see Fig.~\ref{rho_13}(d)) shows that the SL
system is antiferromagnetic at low temperatures. Resistivity curve, plotted in same
figure, shows that the $T_{MIT}$ matches well with the $T_N$. Expectedly the
resistivity of 1/3 SL is much lower than the 1/1 SL due to participations of
thicker uncorrelated layer (see Fig.~\ref{rho_13}(d)).

Does the 1/5 SL system have an antiferromagnetic ordering at low temperatures?
Our calculations show that the induced moment in uncorrelated layers for 1/5 SL
is negligible (see Fig.~\ref{rho_13} (b) and (c)) and as result the long range
antiferromagnetic correlation between the correlated layers cease to exist. To gain
more insight of magnetic correlations of correlated planes separated by five
uncorrelated planes we plot the magnetic structure factor S($\pi$,$\pi$) for individual
correlated planes in the inset of Fig.~\ref{rho_13}(c). It is apparent that the
individual correlated layers are antiferromagnetically ordered by themselves but due to
insufficient induced moments in uncorrelated planes the long range antiferromagnetic
order between them is not established.

\section{$x-T$ phase diagram}

Thereafter, we present the $x-T$ phase diagram for the correlated/uncorrelated
SLs for $U = 8$ where $w_U$ = 1 or/and $w_0$ = 1. The Neel temperature $T_{N}$
[from $S(\pi, \pi, \pi)-T$ plots] and the metal-insulator transition temperature
$T_{MIT}$ [from $\rho_{z}-T$ plot] for different SLs are used to construct this
phase diagram. Fraction of correlated planes $x$ is varied in our SLs. We plot
the $T_N$ for 1/5 SL ($x \sim 0.17$), 1/3 SL ($x \sim 0.25$), 1/1 SL ($x \sim 0.5$),
3/1 SL ($x \sim 0.75$), 5/1 SL ($x \sim 0.83$) and bulk ($x \sim 1$) in
Fig.~\ref{pd_u8}. For bulk case ($x = 1$) the system transits from PM-I phase to
an AF-I phase as we discussed earlier. This PM-I to AF-I transition remains intact
for 5/1 SL ($x \sim 0.83$) although the $T_{MIT}$ decreases considerably. For
$x \le 0.75$ the PM-I phase above AF-I phase disappears and as a result the
$T_{MIT}$ matches well with the $T_N$.
Lastly, the antiferromagnetic ordering vanishes for 1/5 SL (i.e. for $x\sim 0.17$).
So, to summarize, the superlattice system directly transits from AF-I state to
$PM$-$M^{*}$ state with increase of temperature for $x \leq 0.75$ whereas the SL
system converts to $PM$-$M^{*}$ state from AF-I state via the PM-I state upon
increasing the temperature for $x > 0.75$ at $U = 8$. This phase diagram is similar
to the very recently reported phase diagram in Ref.~\citenum{Chakraborty1}, where
randomly diluted systems were studied.

\section{Magnetotransport properties of 3/3 SL}

\begin{figure}[!t]
\centerline{
\includegraphics[width=8.5cm,height=6.3cm,clip=true]{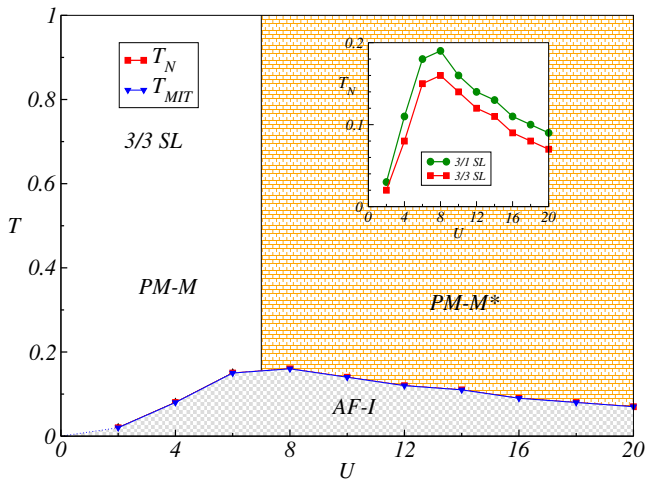}}
\caption{
$U-T$ phase diagram of the 3/3 superlattice. The phase diagram is very similar to
the phase diagram for 3/1 SL in Fig.~\ref{pd_sl}(d). The only difference is that the 
value of $T_{N}$ is suppressed. We compare the $T_N$ values in the inset.
}
\label{pd_33}
\end{figure}

\begin{figure}[!t]
\centerline{
\includegraphics[width=8.5cm,height=6.30cm,clip=true]{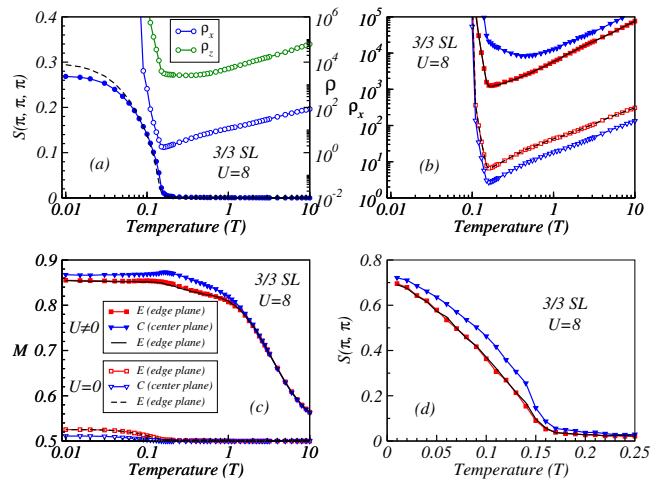}}
\caption{
(a) Temperature evolution of the structure factor (left axis) and out-of-plane
$\rho_{z}$ and in-plane $\rho_{x}$ resistivities (right axis) are plotted for
3/3 SL for $U = 8$. It is apparent that the $T_{MIT}$s obtained from $\rho_{x}$
and $\rho_{z}$ are equal to each other and matches well with the $T_{N}$ of
the SL. The structure factor of 1/1 SL is plotted by dashed line to compare the
$T_N$s. The magnitude of $\rho_{z}$ is much larger than $\rho_{x}$, as expected.
(b) The in-plane resistivity $\rho_{x}$ of the constituent planes of the correlated
($U \ne 0$) and uncorrelated ($U = 0$) layers: The $T_{MIT}$ and the value of
$\rho_{x}$ in the central correlated plane are larger than that of the edge
correlated planes (i.e. $T_{MIT}^{E} < T_{MIT}^{C}$ and $\rho_{x}^{E} < \rho_{x}^{C}$).
Edge correlated planes show MIT exactly at $T_{N}$. All the uncorrelated planes
also show MIT at the same temperature $T = T_{N}$. But, the value of $\rho_{x}$
in the edge uncorrelated plane is somewhat larger than the center uncorrelated
plane ($\rho_{x}^{E} > \rho_{x}^{C}$). The local moment profile of all the
individual planes are presented in (c). In the correlated layer, the local
moments in the edge plane $M_{U}^{E}$ is smaller than the center plane
$M_{U}^{C}$ (i.e. $M_{U}^{E} < M_{U}^{C}$) at low temperatures. Conversely,
in the uncorrelated layer, the induced moments in the edge plane $M_{0}^{E}$
is larger than the center plane $M_{0}^{C}$ (i.e. $M_{0}^{E} > M_{0}^{C}$).
(d) Center correlated plane shows antiferromagnetic 
transition at slightly higher temperature than that of the edge correlated
planes. The value of structure factor in the center plane is also larger
than the edge planes ($S_{U}^{E} < S_{U}^{C}$) at low temperatures.
This is consistent with the moment profile curves shown in (c). Legends are
same in (b), (c), and (d). Physical observables like moment profile,
resistivity and structure factors are symmetric around the central plane
in both correlated and uncorrelated layers.
} 
\label{rho_33}
\end{figure}

Now we proceed to study the plane resolved magnetic and transport properties for
3/3 SL ($x$ = 0.5). First, we plot the $U-T$ phase diagram for 3/3 SL in Fig.~\ref{pd_33}.
This phase diagram is very similar to 3/1 and 1/1 SLs that we presented in
Fig.~\ref{pd_sl}(d). We plot the antiferromagnetic structure factor and resistivities
($\rho_x$ and $\rho_z$) for the 3/3 SL in Fig.~\ref{rho_33}(a) for $U = 8$. The $T_N$
of 3/3 SL remains more or less same to that of 1/1 SL. The $T_{MIT}$ obtained from
$\rho_z$ and $\rho_x$ are also equal to each other and matches well with the $T_N$. 
Obviously, the value of out-of-plane resistivity $\rho_{z}$ is larger than
in-plane resistivity $\rho_{x}$. The plane resolved resistivities of the correlated
and uncorrelated layers are shown in Fig.~\ref{rho_33}(b). The $T_{MIT}$ of
central correlated plane is larger than the edge correlated planes. Interestingly,
only the $T_{MIT}$ of the edge correlated plane matches with the $T_N$. On the
other hand, all the uncorrelated planes show the metal-insulator transition at $T_N$.
In addition it is clear that the central correlated (uncorrelated) plane is more
(less) resistive than the edge correlated (uncorrelated) planes in correlated
(uncorrelated) layer. The coupling of the correlated edge plane with the
nearest uncorrelated plane reduces its resistivity as compared to the central
correlated plane. In other words metallicity penetrates to the correlated edge
layer above the $T_N$ due to the interfacial coupling between the correlated and
uncorrelated layers. This interfacial coupling is also reflected in the plane
resolved magnetic moments profile of the correlated layer shown in Fig.~\ref{rho_33}(c).
We find that the center plane has larger moment as compared to the edge planes
at low temperatures. Expectedly the edge plane of the uncorrelated layer also
has comparatively larger induced magnetic moment to that of center plane as shown
in the same figure.

%***************************************************************************
%***************************************************************************

%***************************************************************************
%***************************************************************************

%***************************************************************************
\begin{figure}[!t]
\centerline{
\includegraphics[width=8.5cm,height=6.30cm,clip=true]{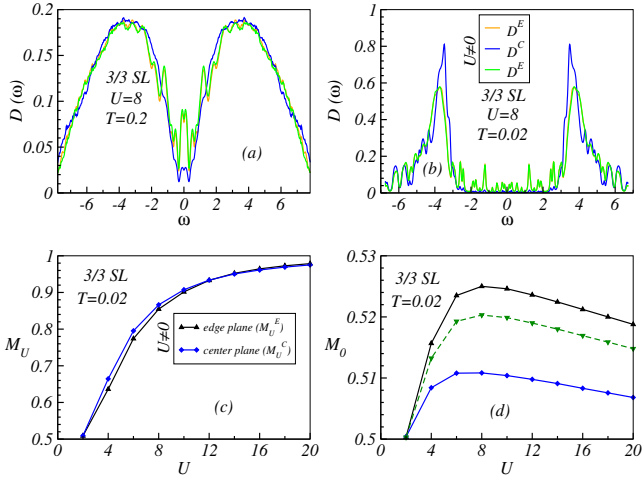}}
\caption{
The density of states (DOS) of the constituent planes of the correlated layer
($U \ne 0$) in the 3/3 SL are plotted for $U = 8$ at (a) $T = 0.2$ (around $T_N$)
and (b) $T = 0.02$ (low temperatures). Mott lobes appear around $\omega = \pm\frac{U}{2}$.
At high temperatures $T = 0.2$ the weight of DOS of the edge correlated plane at
the Fermi level ($\omega = 0$) is greater than that of the center correlated plane
($D^{E} > D^{C}$), supporting the in-plane resistivity calculations shown in
Fig.~\ref{rho_33}(b). A clear Mott gap opens up in both the center and edge planes
of correlated layer at low temperatures $T = 0.02$. But, extra tiny satellite
pattern arises in the band-gap of the edge plane. (c) At low temperatures
($T = 0.02$), the local moment in both the edge and center planes of the correlated
layer increase monotonically with increase of $U$ and saturates at higher $U$ values.
The local moment in the edge plane is smaller than the center plane
($M_{U}^{E} < M_{U}^{C}$) for intermediate values of $U$.
(d) The $U$ dependence of the induced moment in the edge and center
uncorrelated planes at low temperatures $T = 0.02$ exhibits nonmonotonic behavior
(optimum value around $U = 8$). The induced moment in the edge uncorrelated
plane is larger than the center uncorrelated plane ($M_{0}^{E} > M_{0}^{C}$).
The average induced moment profile is plotted by dashed line for comparison.
Legends are same in (c) and (d).
} 
\label{M_33}
\end{figure}
%***************************************************************************

Now, the same question that we addressed for 3/1 SL arises here: Does all the
three correlated planes (two edge planes and one center plane) that constitute the
correlated layer in the 3/3 SL align antiferromagntically at the same temperature?
To answer this question we plot the antiferromagnetic structure factor of
the three correlated planes for $U = 8$ in Fig.~\ref{rho_33}(d). The $T_N$
of the center plane is slightly higher than the edge planes which is expected.
So, one can firmly claim that the effective $U$ ($U_{eff}$) of the central plane
is somewhat larger than the edge planes.

To analyze the plane resolved magnetic and transport properties further we have
calculated the density of states (DOS) of the individual correlated planes of the
3/3 SL. The DOS ($D(\omega)$) of the individual planes of the correlated layer are
shown in Fig.~\ref{M_33}(a) for $U = 8$ at temperature $T = 0.2$ (just above $T_{N}$).
Emergence of Mott lobes at $\omega = \pm\frac{U}{2}$ is apparent in all the three
correlated planes. But the weight of DOS at the Fermi level ($\omega = 0$)
for the edge correlated plane is comparatively larger than to that of the central
correlated plane. This larger weight at the Fermi level enforces the edge correlated
planes to have smaller resistivity than the center correlated plane. At low temperature
$T = 0.02$ a clear Mott gap opens in the DOS of center correlated plane as shown in
Fig.~\ref{M_33}(b). Although same kind of gap is noticed for both the edge plane
and center plane, some very tiny satellite patterned structures appear in the band-gap
of edge planes.

At low temperatures ($T = 0.02$) the local moment of the correlated planes increases
monotonically with increase of $U$ and saturates at large $U$ values as shown in
Fig.~\ref{M_33}(c). The local moments in the center plane is slightly larger than
the edge plane for intermediate $U$ values. Otherwise, the qualitative
monotonic nature of the local moments profile in the correlated planes remains the same.
On the other hand the induced magnetic moment in uncorrelated layer shows nonmonotonic
behavior (for both edge and center plane) with $U$ at low temperatures. The induced
local moment in the edge uncorrelated plane is significantly larger than the center
uncorrelated plane as shown in Fig.~\ref{M_33}(d). This is due to the fact that the
uncorrelated edge plane which is adjacent to the correlated plane gets more affected
by the correlated layer.

\section{Conclusions}
In this paper we have implemented one band Hubbard model at half-filling
to investigate the AF/PM superlattices by using semi-classical Monte Carlo
approach. We analyze various superlattices in three dimensions where correlated
(with on-site repulsion strength $U \ne 0$) and uncorrelated ($U$ = 0) layers
are arranged periodically. First, we explore the $U-T$ phase diagrams for various
$w_U$/1 SLs ($w_U$ = 5, 3, 1) and compare the results with the bulk systems.
We show that the magnetic moments are induced in the uncorrelated layers at low
temperatures due to the kinetic hopping of carriers and optimum magnetic moment
is induced for $U =8$. 
The antiferromagnetic ordering among the induced moments in uncorrelated layers
mediates the long range antiferromagnetic ordering between the correlated layers
and as a result the antiferromagnetic insulating nature of the bulk systems
remains intact in the SLs. In other words, the long range antiferromagnetic order
survives in the superlattices through the mutual cooperation of the induced moment
in the uncorrelated layer and the local moment in the correlated layer. Thus,
the induced moments in the uncorrelated layer play an important role in
determining the global long range antiferromagnetic order in the superlattices.

To analyze the plane resolved magnetotransport properties we focus on the $U = 8$
regime. Interestingly, the average local moments in the edge planes are
comparatively smaller than the central plane in the correlated layer.
The coupling of the edge planes of correlated and uncorrelated layers reduces the
local magnetic moments of the edge correlated plane but induces magnetic moments
in uncorrelated layer. In-plane resistivity calculations of the individual
constituent planes of the superlattices show that the $T_{MIT}$ increases as we move
from edge plane to center plane inside the correlated layers, which is consistent
with the local moment profile of the individual constituent planes. So overall our
plane resolved calculations establish an one-to-one connection between the local
moment profile and the in-plane transport properties of the superlattices.
Plane-resolved density of states calculations of the individual planes of the
correlated layer are also concomitant with the in-plane resistivities.
On the other hand the induced moments in uncorrelated planes decreases considerably
as a function of the distance from the interface, but the metal-insulator transition
temperature of edge and center planes remain more or less unaffected. In the end,
we show that the induced moments in uncorrelated layers dissipates with increasing
the thickness of uncorrelated layers as a result the antiferromagnetic
ordering among correlated layers vanishes.

\begin{center}
\textbf{ACKNOWLEDGMENT}
\end{center}

We acknowledge use of the Meghnad2019 computer cluster at SINP.

\end{document}